\documentclass[conference,a4paper]{APSIPA2026}
\usepackage{amsmath,amssymb,amsfonts}
\usepackage[ruled,vlined]{algorithm2e}
\usepackage{graphicx}
\usepackage{textcomp}
\usepackage{xcolor}
\usepackage{bm}
\usepackage{eucal}
\usepackage{hyperref}
\usepackage{multirow}
\usepackage{array}
\usepackage{booktabs}
\usepackage{pifont}
\usepackage{array}
\usepackage[caption=false,subrefformat=parens,labelformat=parens]{subfig}
\usepackage[backend=biber,style=ieee,]{biblatex}
\usepackage{geometry}
\usepackage{fancyhdr}

\fancypagestyle{firststyle}{
  \fancyhf{}
  \fancyhead[C]{2026 Asia Pacific Signal and Information Processing Association Annual Summit and Conference (APSIPA ASC)}
}

\begin{document}

\title{Energy-Efficient Waveform Design for ISAC Systems: An Ambiguity-Domain QoS Perspective}

\author{
\authorblockN{
Ngoc-Son Duong\authorrefmark{1},
Trung-Hieu Nguyen$^\#$\authorrefmark{1}, and
Quang-Truong Can\authorrefmark{2}
}

\authorblockA{
\authorrefmark{1}
Faculty of Electronic Engineering, Posts and Telecommunications Institute of Technology, Hanoi, Vietnam \\
\authorrefmark{2}
Faculty of Electronics and Telecommunications, VNU University of Engineering and Technology, Hanoi, Vietnam \\
E-mail: \{sondn, hieunt\}@ptit.edu.vn, truongcq@vnu.edu.vn ($^{\#}$Corresponding Author)}
}

\maketitle
\thispagestyle{firststyle}
\pagestyle{empty}

\begin{abstract}
Integrated sensing and communication (ISAC) requires transmit waveforms that simultaneously preserve communication quality, provide reliable sensing, and remain compatible with practical radio-frequency front ends. This paper considers a discrete-time ISAC waveform design problem that minimizes transmit power from the perspective of a novel metric termed the ambiguity-domain sensing signal-to-interference-plus-noise ratio (AF-SINR). The proposed AF-SINR quantifies the ratio between the desired ambiguity-function mainlobe power and the weighted aggregate sidelobe leakage plus noise within a local delay-Doppler region of interest, thereby providing a localized and noise-aware sensing-QoS measure. To enable ISAC operation, the waveform is further required to satisfy per-user effective communication-SINR constraints, while a peak-to-average power ratio (PAPR) constraint is imposed to facilitate practical implementation. The resulting energy-minimization problem is nonconvex due to the fractional QoS expressions, quartic ambiguity terms, and waveform-dependent PAPR constraint. To address this challenge, we propose a fractional-programming successive-convex-approximation (FP-SCA) algorithm. Simulation results verify that the proposed method satisfies all requirements while preserving localized ambiguity suppression over the local delay-Doppler region.
\end{abstract}

\begin{IEEEkeywords}
6G, ISAC, ambiguity-domain sensing, energy-efficient waveform design, fractional programming, successive convex approximation.
\end{IEEEkeywords}

\section{Introduction}
Integrated sensing and communication (ISAC) is regarded as an emerging technology for beyond-5G and sixth-generation wireless networks as it can support data transmission and environmental sensing simultaneously on a common waveform and radio-frequency front end \cite{Liu2022ISAC, Zhang2022JCAS, Liu2020JRC}. In contrast with a communication-only design, an ISAC transmitter must control not only user-level communication quality but also the temporal, spectral, and spatial structure of the probing waveform. This is especially important for orthogonal frequency-division multiplexing (OFDM) systems because their multicarrier structure is well suited for range and Doppler processing. However, the random communication symbols can create high sidelobes in the ambiguity function \cite{Li2025RD,Liu2020JRC}. To handle this problem, a substantial body of research has examined OFDM waveform design from several related perspectives. From an information theoretic viewpoint, Wei \emph{et al.} design ISAC waveform throught power allocation and beamforming problem under sensing and communication mutual information constraints \cite{Wei2024MI}. Meanwhile, He \emph{et al.} improved sensing performance without degrading the communication link by transmitting additional sensing signals through the null space of the communication channel \cite{He2024Null}. Beyond these signal level designs, Zhang \emph{et al.} exploited the flexibility of OFDM across the time and frequency domains by jointly adjusting resource allocation to improve sensing and positioning while preserving communication performance \cite{Zhang2024Cross}. In different scene, practical transmitter limitations have motivated another line of research on PAPR reduction, since large OFDM peaks require greater power amplifier backoff and reduce transmitter efficiency \cite{Huang2022LowPAPR, Varshney2023LowPAPR}. More recently, attention has shifted toward direct control of the ambiguity response. In \cite{Li2025RD}, Li \emph{et al.} minimized the range Doppler integrated sidelobe level under symbol level precoding and constant modulus constraints . Similarly, Feng \emph{et al.} considered weighted peak and integrated sidelobe metrics together with PAPR control and constructive interference constraints for communication \cite{Feng2025WPISL}.

From the above, three observations motivate this work. First, minimizing an integrated sidelobe level (ISL) alone does not define a sensing quality of services (QoS) in the presence of receiver noise, because both the mainlobe and sidelobes scale with waveform energy. Second, while peak-aware or integrated sidelobe metrics are useful, they do not explicitly indicate if the sensing output achieves a required balance between the desired signal (target) and the combined effects of leakage and noise. Third, a practical energy-efficient transmitter should minimize power while satisfying sensing QoS, communication QoS requirements, and a low-PAPR condition. These considerations motivate an ambiguity-domain sensing SINR, denoted AF-SINR, that incorporates the desired ambiguity mainlobe, weighted delay-Doppler sidelobe leakage, and matched-filter noise in one QoS constraint. The main contributions are summarized as follows.
\begin{itemize}
    \item We formulate a transmit-power minimization problem with a novel AF-SINR sensing QoS constraint, per-user communication SINR constraints, and a PAPR constraint. The sensing metric has a direct physical interpretation as the matched-filter mainlobe-to-leakage-plus-noise ratio.
    \item  We develop an FP-SCA algorithm tailored to the proposed ISAC waveform design. The method combines quadratic transforms for the fractional sensing and communication QoS constraints with first-order affine approximations of the ambiguity-function, such that each successive convex subproblem is formulated as a second-order cone program.
\end{itemize}
The remainder of this paper is organized as follows. Section~II introduces the discrete-time ISAC waveform model, the ambiguity-domain sensing metric, and the minimum-energy waveform design problem under communication-QoS and PAPR constraints. Section~III presents the proposed FP-SCA algorithm. Section~IV provides numerical results and comparisons with benchmark designs. Finally, Section~V concludes the paper and discusses future research directions.
\section{Problem Formulation}
\subsection{Ambiguity-Domain sensing SINR}
Consider a complex baseband transmit block represented by
\begin{equation}
\mathbf{x} = [x_0,x_1,\ldots,x_{N-1}]^\top \in \mathbb{C}^{N},
\end{equation}
where $N$ is the discrete waveform length. The vector can represent a time-domain OFDM block or a vectorized waveform after fixing a transmit architecture. Let $\tau$ and $\nu$ denote integer circular delay and normalized Doppler bins, respectively. The circular delay matrix $\mathbf{J}_{\tau}$ and Doppler modulation matrix $\mathbf{D}_{\nu}$ are defined such that
\begin{equation}
[\mathbf{J}_{\tau}\mathbf{x}]_n=x_{(n-\tau)\bmod N},
\end{equation}
and
\begin{equation}
\mathbf{D}_{\nu}=\operatorname{diag}\left(e^{j2\pi\nu n/N}\right)_{n=0}^{N-1}
\end{equation}
The discrete periodic ambiguity sample at the delay-Doppler cell $(\tau, \nu)$ is
\begin{equation}\label{eq:af_sample}
\chi_{\tau,\nu}(\mathbf{x}) = \mathbf{x}^{\mathsf{H}}\mathbf{D}_{\nu}\mathbf{J}_{\tau}\mathbf{x}
 = \mathbf{x}^{\mathsf{H}}\mathbf{Q}_{\tau,\nu}\mathbf{x}.
\end{equation}
Let the origin cell be the ambiguity mainlobe. Since $\mathbf{Q}_{0,0} = \mathbf{I}$, its value is
\begin{equation}\label{eq:mainlobe}
\chi_{0,0}(\mathbf{x})=\|\mathbf{x}\|_2^2.
\end{equation}
Let $\CMcal{Q}$ be a prescribed region of interest (RoI) excluding the origin. This region identifies delay-Doppler cells where sidelobes can mask weak echoes or create false detections \cite{Zhang2024Cross}. We define the AF-SINR as
\begin{equation}\label{eq:afsinr}
\Gamma_{\mathrm{AF}}(\mathbf{x})=
\frac{|\chi_{0,0}(\mathbf{x})|^2}{\sum_{q\in\CMcal{Q}}w_q|\chi_q(\mathbf{x})|^2+
\sigma_{\mathrm{AF}}^2\|\mathbf{x}\|_2^2+\sigma_0^2}.
\end{equation}
where $w_q\geq0$ denotes the weight assigned to the $q$-th delay-Doppler cell. The numerator in \eqref{eq:afsinr} is the desired ambiguity mainlobe power. The first term in the denominator is the weighted sidelobe leakage over the RoI. The second term models the output noise power of a matched filter whose reference waveform scales with the transmit block, while the third represents a waveform-independent noise. Physically, it measures the ability of the matched-filter response to echo from nearby clutter, neighboring targets, and noise. Unlike the integrated sidelobe level \cite{Li2025RD}, which only measures the total sidelobe energy, the AF-SINR explicitly accounts for the mainlobe strength and receiver noise. Similarly, whereas the peak sidelobe level \cite{Zhang2024Cross} focuses only on the strongest sidelobe, the AF-SINR captures the cumulative effect of all weighted sidelobes in the local RoI. The metric can also be interpreted as a counterpart of the signal-to-clutter-plus-noise ratio, although it does not require explicit target or clutter coefficients. This makes AF-SINR particularly suitable for waveform optimization, since it directly imposes a sensing-QoS threshold while jointly reflecting local ambiguity suppression, transmit energy, and noise robustness. Similar to standard SINR requirements, we need AF-SINR satisfies 
\begin{equation}
\Gamma_{\mathrm{AF}}(\mathbf{x})\geq\gamma_{\mathrm{AF}},
\label{eq:af_qos}
\end{equation}
where $\gamma_{\mathrm{AF}}$ is a required AF-SINR threshold.

\subsection{Communication QoS}
\label{subsec:communication_model}

In addition to the sensing functionality, the transmit waveform is required to support downlink communication with $K$ users. Rather than explicitly modeling a conventional multiuser precoder and symbol vector, we adopt an effective waveform-response model. For $k$-th user, let $\mathbf{d}_k\in\mathbb{C}^{N}$ denote the desired waveform-response vector, where the scalar projection $\mathbf{d}_k^{\mathsf{H}}\mathbf{x}$ characterizes the useful communication component induced by $\mathbf{x}$. Moreover, let $\mathbf{I}_k= \left[ \mathbf{i}_{k,1}, \mathbf{i}_{k,2}, \ldots, \mathbf{i}_{k,L_k} \right] \in\mathbb{C}^{N\times L_k}$ collect the interference-response vectors associated with user $k$. The interference term $\mathbf{I}_k^{\mathsf{H}}\mathbf{x}$ therefore aggregates the waveform-dependent residual interference across $L_k$ effective interference components. Accordingly, the desired signal power and the interference power experienced by $k$-th user are respectively given by
\begin{equation}
    S_k(\mathbf{x}) = \left| \mathbf{d}_k^{\mathsf{H}}\mathbf{x} \right|^2,
\end{equation}
and
\begin{equation}
    I_k(\mathbf{x}) = \left\| \mathbf{I}_k^{\mathsf{H}}\mathbf{x} \right\|_2^2 = \sum_{\ell=1}^{L_k} \left| \mathbf{i}_{k,\ell}^{\mathsf{H}}\mathbf{x} \right|^2.
\end{equation}
Let $\sigma_{\mathrm{C},k}^2$ denote the effective noise power at $k$-th user, which may include thermal noise. The resulting communication SINR is modeled as
\begin{equation}
\Gamma_k(\mathbf{x})=
\frac{ \left| \mathbf{d}_k^{\mathsf{H}}\mathbf{x}
\right|^2 }{ \left\| \mathbf{I}_k^{\mathsf{H}}\mathbf{x}
\right\|_2^2 + \sigma_{\mathrm{C},k}^2 }, \quad k=1,\ldots,K.
\label{eq:comm_sinr}
\end{equation}
It is noted that \eqref{eq:comm_sinr} is a reduced form of the linear communication SINR. Specifically, the channel coefficients, fixed transmit/receive beamformers, and linear waveform-projection operators are absorbed into the effective desired-response vector $\mathbf{d}_k$ and the interference-response matrix $\mathbf{I}_k$. These quantities are assumed to be known, e.g., from channel estimation and predetermined transceiver processing. To guarantee the communication QoS, the waveform must satisfy
\begin{equation}
\Gamma_k(\mathbf{x}) \geq \gamma_k, \quad k=1,\ldots,K,
\label{eq:comm_qos}
\end{equation}
where $\gamma_k>0$ is the prescribed SINR target.




\subsection{PAPR constraint}
Apart from AF-SINR and communication QoS, another important requirement is the low PAPR. Maintaining a low PAPR is critical for maximizing both hardware efficiency and overall communication and sensing performance. A waveform with a low PAPR allows the transmitter's power amplifier to operate closer to its saturation region, which significantly enhances energy efficiency without inducing severe non-linear distortion. The discrete-time PAPR is defined as
\begin{equation}
\operatorname{PAPR}(\mathbf{x})=
\frac{N\max_n|x_n|^2}{\|\mathbf{x}\|_2^2}.
\label{eq:papr}
\end{equation}
A finite upper bound $\rho>1$ is imposed as
\begin{equation}
|x_n|^2\leq\frac{\rho}{N}\|\mathbf{x}\|_2^2,
\quad n = 0,...,N-1.
\label{eq:papr_constraint}
\end{equation}
Herein, parameter $\rho$ balances envelope regularity and design flexibility. A smaller $\rho$ improves hardware compatibility but restricts amplitude degrees of freedom. The limiting case $\rho=1$ corresponds to constant-modulus (CM) signaling. In this work, we prefer a design with low PAPR because the feasible of CM set degenerates into a nonconvex phase manifold. We refer the reader to manifold optimization \cite{manopt} or gradient projection methods \cite{gp} if interested in CM constraints.

\subsection{Problem formulation}
Along with the constraints mentioned above, the waveform energy minimization problem is formulated as
\begin{subequations} \label{eq:problem_original}
\begin{align}
\underset{\mathbf{x}\in\mathbb{C}^{N}}{\mathsf{minimize}}
\quad & \|\mathbf{x}\|_2^2 \label{eq:objective} \\
\mathsf{subject~to}
\quad & \Gamma_{\mathrm{AF}}(\mathbf{x})\geq\gamma_{\mathrm{AF}}, \label{eq:const_af} \\
& \Gamma_k(\mathbf{x})\geq\gamma_k,
\quad k=1,\ldots,K, \label{eq:const_k} \\
& |x_n|^2\leq\frac{\rho}{N}\|\mathbf{x}\|_2^2,
\quad n=0,\ldots,N-1. \label{eq:const_power}
\end{align}
\end{subequations}
Problem \eqref{eq:problem_original} is nonconvex mainly because the AF-SINR contains quartic terms. In the next section, we  develop a conic method for obtaining a feasible solution.

\section{Proposed FP-SCA Method}
\subsection{Quadratic transform for QoS ratios}
At $r$-th iteration, using Theorem 2 in \cite{Shen2018FP}, we update the sensing auxiliary variable as
\begin{equation}
y_{\mathrm{AF}}^{(r)}=
\frac{\chi_{0,0}(\mathbf{x}^{(r)})}
{\sum_{q\in\CMcal{Q}}w_q|\chi_q(\mathbf{x}^{(r)})|^2+
\sigma_{\mathrm{AF}}^2\|\mathbf{x}^{(r)}\|_2^2+\sigma_0^2}.
\label{eq:yaf_update}
\end{equation}
Similarly, the $k$-th communication auxiliary variable is
\begin{equation}
y_k^{(r)}=
\frac{\mathbf{d}_k^{\mathsf{H}}\mathbf{x}^{(r)}}
{\|\mathbf{I}_k^{\mathsf{H}}\mathbf{x}^{(r)}\|_2^2+\sigma_{\mathrm{C},k}^2}.
\label{eq:yk_update}
\end{equation}
These updates convert the fractional QoS requirements into difference-of-quadratic expressions when the auxiliary variables are fixed \cite{Shen2018FP}.

\subsection{Local ambiguity approximation}
The ambiguity sample in \eqref{eq:af_sample} is a complex quadratic form. At $\mathbf{x}^{(r)}$, we use the local model
\begin{equation}
\hat{\chi}_q(\mathbf{x};\mathbf{x}^{(r)})=
(\mathbf{x}^{(r)})^{\mathsf{H}}\mathbf{Q}_q\mathbf{x}
+\mathbf{x}^{\mathsf{H}}\mathbf{Q}_q\mathbf{x}^{(r)}
-(\mathbf{x}^{(r)})^{\mathsf{H}}\mathbf{Q}_q\mathbf{x}^{(r)}.
\label{eq:chi_local}
\end{equation}
It creates an affine complex expression in the variable vector and enables a conic representation of the aggregate sidelobe term. Because general delay-Doppler operators are not necessarily Hermitian, \eqref{eq:chi_local} is employed as a local sequential approximation. Consequently, the original QoS constraints are explicitly rechecked after each update, and a restoration step is applied before accepting the next iterate. For the mainlobe, the local approximation reduces to
\begin{equation}
\hat{\chi}_{0,0}(\mathbf{x};\mathbf{x}^{(r)})=
2\operatorname{Re}\{\left(\mathbf{x}^{(r)}\right)^{\mathsf{H}}\mathbf{x}\}
-\|\mathbf{x}^{(r)}\|_2^2.
\label{eq:mainlobe_local}
\end{equation}
With fixed $y_{\mathrm{AF}}^{(r)}$, the sensing QoS is approximated as
\begin{equation}
\begin{split}
&\gamma_{\mathrm{AF}} +|y_{\mathrm{AF}}^{(r)}|^2
\bigg( \sum_{q\in\CMcal{Q}}w_q |\hat{\chi}_q(\mathbf{x};\mathbf{x}^{(r)})|^2 \\
& +\sigma_{\mathrm{AF}}^2\|\mathbf{x}\|_2^2+\sigma_0^2 \bigg)
\leq 2\operatorname{Re}\left\{
\left(y_{\mathrm{AF}}^{(r)}\right)^* \hat{\chi}_{0,0}(\mathbf{x};\mathbf{x}^{(r)})\right\}.
\end{split}
\label{eq:af_surrogate}
\end{equation}
Similarly, the communication QoSs become
\begin{align}\label{eq:comm_surrogate}
\gamma_k+|y_k^{(r)}|^2
\left( \|\mathbf{I}_k^{\mathsf{H}}\mathbf{x}\|_2^2+\sigma_{\mathrm{C},k}^2 \right) \leq 2\operatorname{Re}\left\{
\left(y_k^{(r)}\right)^*\mathbf{d}_k^{\mathsf{H}}\mathbf{x} \right\}.
\end{align}

\subsection{PAPR successive convex approximation}
The right-hand side of \eqref{eq:papr_constraint} contains $\|\mathbf{x}\|_2^2$, which is convex and therefore cannot be retained directly on the right-hand side of a convex inequality. Its affine lower bound at $\mathbf{x}^{(r)}$ is
\begin{equation}
\underline{p}(\mathbf{x};\mathbf{x}^{(r)})=
2\operatorname{Re}\{\left(\mathbf{x}^{(r)}\right)^{\mathsf{H}}\mathbf{x}\}
-\|\mathbf{x}^{(r)}\|_2^2.
\label{eq:power_lower_bound}
\end{equation}
Then, we can use a PAPR surrogate, which is defined as
\begin{equation}
|x_n|^2\leq
\frac{\rho}{N}\underline{p}(\mathbf{x};\mathbf{x}^{(r)}),
\quad n=0,\ldots,N-1.
\label{eq:papr_surrogate}
\end{equation}

\subsection{Convexified problem}
After convexing the cost function and constraints, problem in \eqref{eq:problem_original} becomes
\begin{equation}\label{cvx_prob}
\begin{aligned} \mathsf{minimize}_{\mathbf{x}}\quad & \|\mathbf{x}\|_2^2 \\
    \mathsf{subject~to}\quad 
    & \eqref{eq:af_surrogate}, \eqref{eq:comm_surrogate}, \eqref{eq:papr_surrogate}.
\end{aligned}
\end{equation}

\subsection{Feasibility restoration}
The local ambiguity model in \eqref{eq:chi_local} is not a global inner approximation for arbitrary non-Hermitian $\mathbf{Q}_q$. To retain an original-QoS feasible waveform, let $\mathbf{x}=\sqrt{P}\mathbf{u}$, $\|\mathbf{u}\|_2=1$, and $B(\mathbf{u}) = \sum_{q\in\CMcal{Q}}w_q
|\mathbf{u}^{\mathsf{H}}\mathbf{Q}_q\mathbf{u}|^2$. The original AF-SINR becomes
\begin{equation}
\Gamma_{\mathrm{AF}}(P,\mathbf{u})=
\frac{P^2}{P^2B(\mathbf{u})+P\sigma_{\mathrm{AF}}^2+\sigma_0^2}.
\label{eq:af_power_shape}
\end{equation}
Therefore, the minimum power is the positive root of
\begin{equation}
\left[1-\gamma_{\mathrm{AF}}B(\mathbf{u})\right]P^2
-\gamma_{\mathrm{AF}}\sigma_{\mathrm{AF}}^2P
-\gamma_{\mathrm{AF}}\sigma_0^2=0,
\label{eq:af_power_root}
\end{equation}
provided that $1-\gamma_{\mathrm{AF}}B(\mathbf{u})>0$. The solution to \eqref{eq:af_power_root} is
\begin{equation}\label{root}
    P_{\rm{AF}}(\mathbf{u}) = \frac{\gamma_{\rm{AF}}\sigma^2_{\rm{AF}} + \sqrt{\gamma_{\rm{AF}}^2\sigma^4_{\rm{AF}} + 4\gamma_{\rm{AF}}\sigma_0^2(1 - \gamma_{\rm{AF}}B(\mathbf{u}))}}{2(1 - \gamma_{\rm{AF}}B(\mathbf{u}))}.
\end{equation}
Likewise, define $S_k(\mathbf{u})=|\mathbf{d}_k^{\mathsf{H}}\mathbf{u}|^2$ and $I_k(\mathbf{u})=\|\mathbf{I}_k^{\mathsf{H}}\mathbf{u}\|_2^2$, the minimum power required for user $k$ is
\begin{equation}
P_{\mathrm{C},k}(\mathbf{u})=
\frac{\gamma_k\sigma_{\mathrm{C},k}^2}
{S_k(\mathbf{u})-\gamma_k I_k(\mathbf{u})},
\label{eq:comm_power_root}
\end{equation}
whenever the denominator is positive. The restored power is
\begin{equation}\label{eq:restoration_power}
P_{\mathrm{req}}(\mathbf{u})=
\max\left\{P_{\mathrm{AF}}(\mathbf{u}),P_{\mathrm{C},1}(\mathbf{u}),\ldots,P_{\mathrm{C},K}(\mathbf{u})\right\}.
\end{equation}
Since PAPR is scale invariant, restoring $\mathbf{x}$ by $\sqrt{P_{\mathrm{req}}}\mathbf{u}$ maintains the candidate PAPR while preserving the original AF-SINR and communication SINR requirements. The pseudocode for our proposed algorithm is given by Algorithm \ref{alg:fp_sca}.
\begin{algorithm}[t]
\caption{Proposed SCA-based algorithm for \eqref{eq:problem_original}}
\label{alg:fp_sca}
\KwIn{
Initial waveform $\mathbf{x}^{(0)}$, maximum iteration
number $r_{\max}$, and tolerance $\epsilon$
}
\KwOut{
$\mathbf{x}^{\star}$
}

$r\leftarrow 0$\;
\While{$r<r_{\max}$}{
Update $y_{\mathrm{AF}}^{(r)}$ and $\{y_k^{(r)}\}_{k=1}^{K}$ using
\eqref{eq:yaf_update} and \eqref{eq:yk_update}\;
Solve \eqref{cvx_prob} and denote its solution by $\tilde{\mathbf{x}}^{(r+1)}$\;

$\mathbf{u}^{(r+1)}\leftarrow
\tilde{\mathbf{x}}^{(r+1)}/ \|\tilde{\mathbf{x}}^{(r+1)}\|_2$\;

Compute $B(\mathbf{u}^{(r+1)})$, $\{S_k(\mathbf{u}^{(r+1)})\}_{k=1}^{K}$, and $\{I_k(\mathbf{u}^{(r+1)})\}_{k=1}^{K}$\;

\If{
$1-\gamma_{\mathrm{AF}}B(\mathbf{u}^{(r+1)})\leq 0$
\textbf{or}
$\exists k\in\{1,\ldots,K\}:
S_k(\mathbf{u}^{(r+1)})-\gamma_k
I_k(\mathbf{u}^{(r+1)})\leq 0$
}{
\textbf{break}
}

Compute $P_{\mathrm{AF}}(\mathbf{u}^{(r+1)})$ from
\eqref{root}\;

Compute $P_{\mathrm{C},k}(\mathbf{u}^{(r+1)})$ from
\eqref{eq:comm_power_root} for all $k$\;

Compute restored power by \eqref{eq:restoration_power}\;

$\mathbf{x}^{(r+1)}\leftarrow
\sqrt{P_{\mathrm{req}}^{(r+1)}}
\mathbf{u}^{(r+1)}$\;

\If{
$\dfrac{
\left|\|\mathbf{x}^{(r+1)}\|_2^2 - \|\mathbf{x}^{(r)}\|_2^2
\right|}{ \max\left\{1,\|\mathbf{x}^{(r)}\|_2^2 \right\}}
\leq\epsilon$}{\textbf{break}\;}
$r\leftarrow r+1$\;}
$\mathbf{x}^{\star}\leftarrow\mathbf{x}^{(r)}$\;
\Return{$\mathbf{x}^{\star}$}\;
\end{algorithm}

\section{Simulation Results and Discussion}
\subsection{Simulation Configuration}
For performance evaluation, the proposed method is compared against a radar-only design, which is formulated by dropping the communication-QoS constraints while maintaining the same AF-SINR and PAPR requirements. The principal performance metric is the required  transmit power, reported in dBm, when \textit{i)} changing the AF-SINR threshold while the communication-QoS target remains fixed and \textit{ii)} changing the communication-QoS requirement while the AF-SINR threshold is held fixed. In this simulation, the ISAC system has to serve $K = 3$ users using a waveform of the length $N = 32$. The reference sensing-QoS target is $\gamma_{\rm AF}=16$ dB, while the communication-SINR target is set to $\gamma_k=20$ dB for all users. Moreover, the PAPR threshold is fixed at $\rho=1.5$, and the maximum number of iterations is set to $10$.
\subsection{Simulation results and Discussion}
\subsubsection{On the convergence behavior}
\begin{figure}[ht]
\centering
    \subfloat[Transmit-power convergence\label{subfig:beam1}]{%
        \includegraphics[width=1\linewidth]{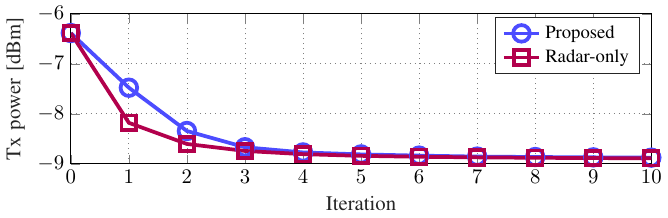}%
    }
    \hfill
    \subfloat[AF-SINR convergence\label{subfig:beam2}]{%
        \includegraphics[width=1\linewidth]{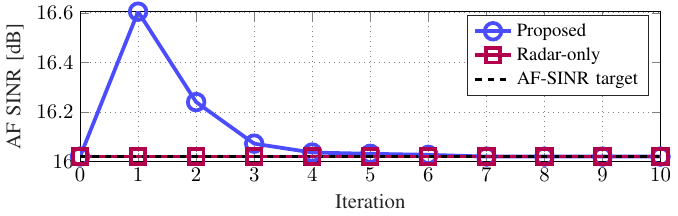}%
    }
    \hfill
    \subfloat[PAPR convergence\label{subfig:beam3}]{%
        \includegraphics[width=1\linewidth]{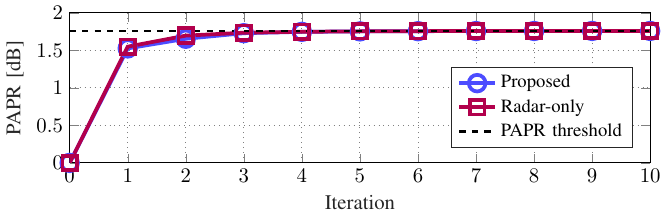}%
    }
    \caption{Convergence behavior of the proposed ISAC and radar-only designs.}\label{fig:convergence}
\end{figure}
Fig.~\ref{fig:convergence} illustrates the convergence behavior of the proposed design and radar-only design, both initialized under same conditions. As can be seen, both methods reduce the required transmit power and reach a stable-state after four iterations. Interestingly, while the radar-only design converges directly to the target AF-SINR, the proposed method fluctuate before stabilizing. Moreover, the PAPR values of both waveforms gradually increase and finally approach the prescribed threshold, indicating that the PAPR constraint becomes active at convergence. Crucially, even with the communication-QoS requirements, the proposed design achieves a same transmit power compared to that of the radar-only scheme.
\subsubsection{Visualizing local ambiguity-domain shaping}
\begin{figure*}
    \centering
    \includegraphics[width=1\linewidth]{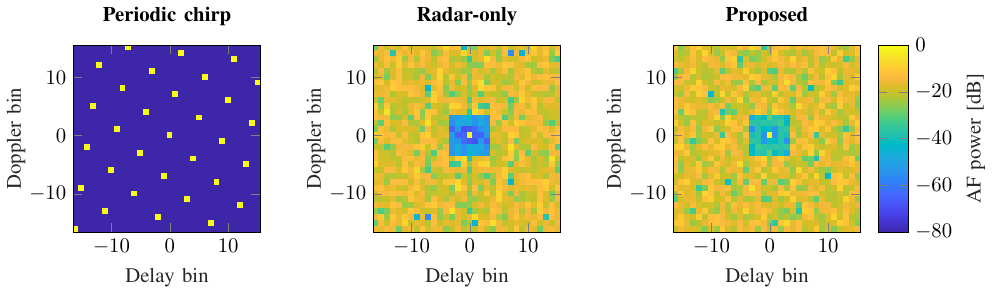}
    \caption{Visual comparison of normalized periodic ambiguity-function power maps for the periodic chirp \cite{lfm}, radar-only, and proposed ISAC waveforms.}
    \label{fig:af}
\end{figure*}
Fig. \ref{fig:af} shows the ambiguity-shaping characteristics among the three waveforms. As can be seen, the periodic chirp exhibits a highly structured ambiguity pattern, with a low background level but several periodic replicas that appear as isolated high peaks across the delay-Doppler plane. Despite its high sensing efficiency, this waveform offers no communication capability. In contrast, the radar-only and proposed waveforms explicitly shape the AF over the local delay-Doppler RoI. The radar-only waveform exhibits the darkest colors within the protected RoI, indicating the strongest local sidelobe suppression because all waveform degrees of freedom are devoted to sensing function. The proposed waveform preserves the same RoI-centered low sidelobe structure, albeit with a slightly brighter RoI than the radar-only design. This increase reflects the sensing-communication trade-off, as part of the waveform design freedom must be allocated to satisfy the communication QoS constraints. Lastly, we can recognize that ambiguity energy outside the RoI is allowed to remain relatively high. This can be exploited to satisfy the communication requirements while preserving the prescribed local sensing performance \cite{Zhang2024Cross}.

\subsubsection{Transmit-power requirement under varying AF-SINR targets}
\begin{figure}
    \centering
    \includegraphics[width=0.8\linewidth]{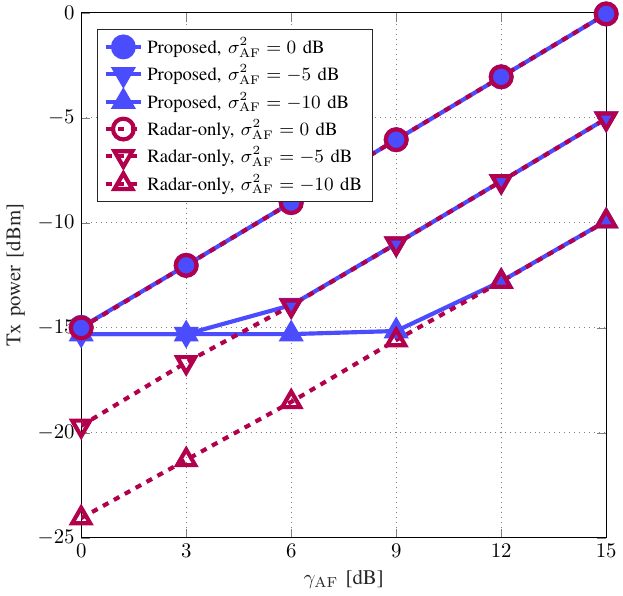}
    \caption{Required transmit power versus AF-SINR requirement.}
    \label{fig:gammaAF_change}
\end{figure}
Fig. \ref{fig:gammaAF_change} depicts the required transmit power as a function of the AF-SINR requirement under different filter noise. As we can see, the transmit power increases monotonically with $\gamma_{\rm AF}$, and a larger $\sigma_{\rm AF}^2$ consistently incurs a higher power cost. Notably, at low $\gamma_{\rm AF}$ for the case $\sigma_{\rm AF}^2 = -5$ dB and $\sigma_{\rm AF}^2 = -10$ dB, the proposed design requires a higher transmit power than the radar-only scheme because the sensing constraint is relatively loose, whereas the communication QoS constraints remain active. In this regime, the minimum transmit power is determined by the communication-induced power requirement rather than by the AF-SINR requirement. As $\gamma_{\rm AF}$ increases, the sensing constraint becomes dominant for both schemes. Consequently, the power gap decreases and the two curves tend to converge.

\subsubsection{Transmit-power requirement under varying communication-QoS targets}
\begin{figure}
    \centering
    \includegraphics[width=0.82\linewidth]{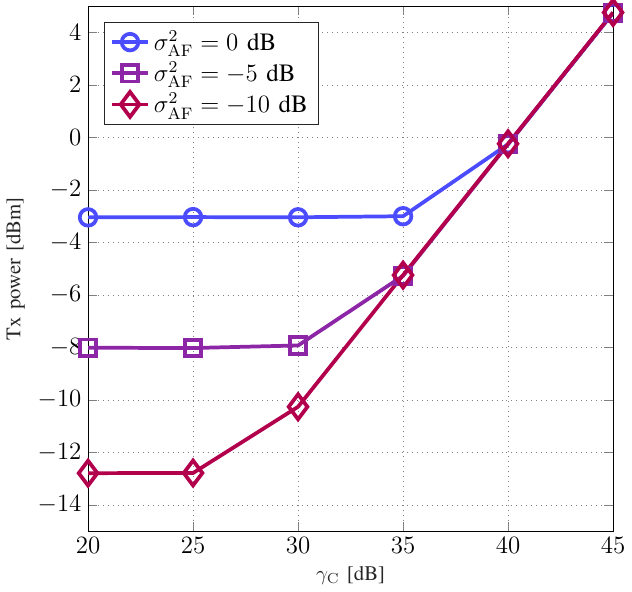}
    \caption{Required transmit power versus communication QoS requirement.}
    \label{fig:comQoS}
\end{figure}
Fig.~\ref{fig:comQoS} illustrates the required transmit power versus the communication QoS target $\gamma_{\rm C}$ under different filter noise coefficients. At low communication-QoS requirements, each curve exhibits an approximately constant power floor. This behavior indicates a sensing-limited regime, in which the AF-SINR constraint dominates the power requirement. Accordingly, a larger matched-filter noise coefficient $\sigma_{\rm AF}^2$ leads to a higher sensing-induced power floor, since additional transmit energy is required to compensate for the noise term in the AF-SINR denominator. As $\gamma_{\rm C}$ increases, the communication QoS constraints gradually become active, causing the required power to increase. The curves associated with smaller $\sigma_{\rm AF}^2$ begin to rise at lower $\gamma_{\rm C}$ values because their sensing-induced power floors are lower. Hence, the communication constraint becomes dominant earlier. At sufficiently stringent communication-QoS targets, the three curves nearly converge, demonstrating that the transmit power is then primarily determined by the communication requirement rather than the sensing noise level. 
\section{Conclusions}
This paper investigated a minimum-energy ISAC waveform design problem under AF-SINR requirement, per-user communication QoS constraints, and a PAPR constraint. The proposed AF-SINR provides a localized sensing-QoS measure by comparing the desired ambiguity mainlobe power against the weighted sidelobe leakage and noise within a local delay-Doppler RoI. To handle the resulting nonconvex problem, an FP-SCA algorithm was developed to iteratively obtain feasible low-energy waveform solutions. Simulation results showed that the proposed waveform preserves localized ambiguity suppression while satisfying all constrains. The results also reveal the trade-off, in which the proposed waveform incurs a communication-induced power penalty under relaxed sensing requirements, whereas its power consumption approaches that of the radar-only design when the AF-SINR requirement becomes dominant. Future work will consider more detailed comparisons with waveform designs based on alternative sensing metrics, such as ISL, PSL, as well as structured waveforms including OFDM, chirp-based, and phase-coded designs.
\printbibliography
\end{document}